# VR-CAD Framework for Parametric Data Modification with a 3D Shape-based Interaction

Yujiro Okuya [1, 2]     Nicolas Ladeveze [1]     Cédric Fleury [2]     Patrick Bourdot [1]

yujiro.okuya@limsi.fr     nicolas.ladeveze@limsi.fr     cedric.fleury@lri.fr     patrick.bourdot@limsi.fr

1. LIMSI/CNRS, Université Paris-Saclay F-91400 Orsay, France
2. LRI, Univ. Paris-Sud, CNRS, Inria, Université Paris-Saclay F-91400 Orsay, France

**Keywords**: Virtual Reality, Computer aided design, Parametric CAD, Haptics.

## 1. Introduction

Most of Virtual Reality (VR) applications for Computer Aided Design (CAD) focused on immersive project reviewing or early prototyping in order to reduce cost and time of product design cycle. In terms of design activity, current VR technique enables industrial designers to draw preliminary drafts or edit simple primitives with their hand gestures in 3D space. Such applications enhance designer's inspiration and improve their spatial thinking for aesthetic design. However, hand gesture based interaction has not yet been applied for parametric design tools due to the complexity of CAD data. Industrial models are necessarily modeled with parametric CAD systems (e.g. CATIA, SolidWorks, Inventor, etc.) based on preliminary drafts by CAD engineers, in order to support detailed design requirements and numerical simulations for product's manufacturability. The different focus of industrial designers and CAD engineers impose them to use distinct tools for design activities. This disparity interferes with their communication in early design stage.

In this poster, we present a new VR-CAD framework, allowing user to modify parametric CAD data with 3D interaction in an immersive environment. With this framework, users can implicitly modify parameter values of CAD data with co-localized 3D shape-based interaction. This poster describes the system architecture and the interaction technique based on it.

## 2. Related Work

Most VR-CAD applications place value on coherent dimension between visualization and interaction space. According to Clark, "To expect a designer of 3-D surfaces to work with 2-D input and viewing devices unnecessarily removes a valuable degree of freedom" [1]. 3D interaction technique has been mainly carried out for drawing tools and surface modeling, or a few of them are applied for manipulation of 3D primitives. Pushing/pulling interaction for object manipulation was introduced in [2], and MockUp Builder [3] applied this interaction for object creation/edition on multi-touch table top display with co-localized manipulation. However, it is difficult to apply these interaction techniques to complex parametric CAD modeling.

A parametric CAD model is defined by a set of operations (Extrusion, Sweep, Boolean, and so on) applied from primitives and 2D sketches, based on a number of parameters and geometrical or topological constraints. In order to update the CAD model, users need to select a specific constraint in Constructive History Graph (CHG) and modify its parameter value. cReaRV [4] adapted this selection method to 3D space by allowing users to implicitly select a parameter value from selection of a relevant surface with a flystick. Nevertheless, parametric modification still remains in one dimension: users increase or decrease a parameter value by a scroll motion of their hands in 3D space. This one dimensional parametric modification can hardly be transformed into 3D interaction, especially with the complex model containing several constraints.

## 3. VR-CAD System Architecture

To address this issue, we designed a new VR-CAD system based on a distributed architecture (Figure 1). It consists of three components: a VR-CAD server, a Visuo-Haptic server (VHServer) and a VR platform. The VR-CAD server interacts with a CAD engine (CATIA V5) to load and update any parameter modifications into the CAD data structure. The VHServer handles user interaction during modification (see section 4), and supports force feedback device with a high communication rate. The VR platform manages the graphic rendering of virtual environment (VE) with Unity. Heterogeneous platforms, such as CAVE, Wall-sized display and HMD, are supported. Figure 2 is an example of this VR-CAD application in a CAVE system with a force feedback device.

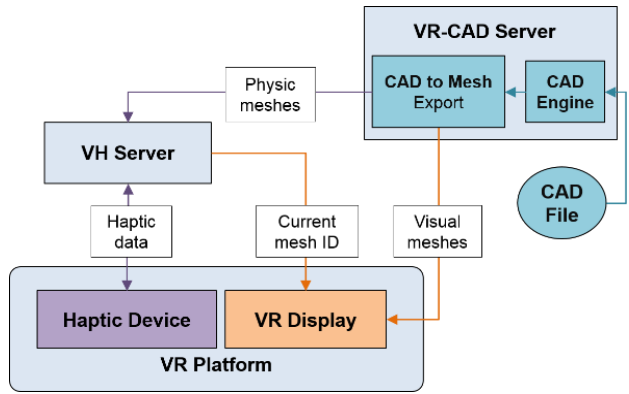
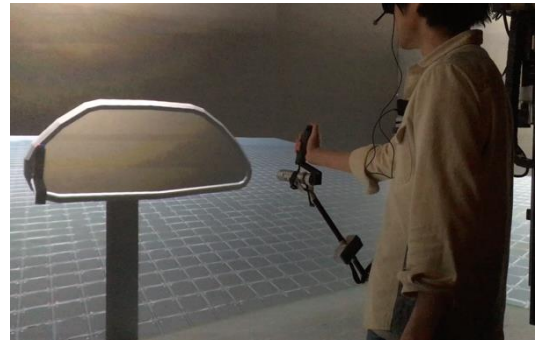

**Figure 1.** System architecture of the distributed VR-CAD system.

**Figure 2.** Example of the application in a CAVE system with a force feedback device.

## 4. 3D Shape-based Interaction to Modify CAD-data Parameters

Main difficulty of a 3D shape-based interaction with a parametric CAD model is the "unpredictable behavior" of shape evolution due to the internal constraints. Industrial CAD model often contains geometrical or topological constraints between two or more entities to define the design rules or limit its physical position and motion.

In order to anticipate the direction of the shape deformation, we compute several meshes by a set of discrete parameter values with a slight offset from the initial value. In such a way, users can choose one of the proposed shape with a 3D hand motion. Only the closest shape from user's hand is rendered in the VE, so that the shape appears to be following user's hand as a pushing/pulling interaction on surface regardless of hidden internal topology complexity.

When a user select a parameter by picking a specific part of a CAD object in the VE, the information about the selected part and a set of new parameter values are sent to the VR-CAD server. Then, the VR-CAD server tessellates different B-Rep data computed from the new parameter set with the CAD engine and save generated shapes (.obj files) into a shared folder. The VH server imports mesh data and computes a distance between meshes and user's hand position. It then transmits the closest mesh ID to a VR platform to switch the visualized mesh.

## 5. Conclusion and Future Work

This paper presents a new VR-CAD framework for CAD-data modification in an immersive environment. The contributions of this framework are to support (i) native CAD-data modification in VR, (ii) 3D shape-based interaction for parametric CAD-data modification, and (iii) CAD design activity over heterogeneous VR systems. We are currently investigating users' assessment of the proposed interaction technique and collaborative design scenario. We expect that the framework improves the communication between industrial designers and CAD engineers in early design stage by providing a common design space.